\documentclass[aps,twocolumn,preprintnumbers,pra]{revtex4-2}
\usepackage{graphicx}  % Include figure files
\usepackage{subfigure}
\usepackage{multirow}
\usepackage{mathtools}
\usepackage{notes2bib}
\Large

\usepackage[utf8]{inputenc}
\usepackage[english]{babel}

\usepackage{fancyhdr}
\usepackage{longtable}
\usepackage{parskip}
\usepackage[T1]{fontenc}
\usepackage{dcolumn}   % Align table columns on decimal point

\usepackage{bm}        % bold math
\usepackage{amsfonts}  % Common math fonts
\usepackage{amsmath}   % Common math functions
\usepackage{amssymb}   % Common math symbols

\newcommand{\pwisein}{\left\{ \begin{array}{ll}}
\newcommand{\pwiseout}{\end{array}\right.}

\usepackage[unicode=true,pdfusetitle,
 bookmarks=true,bookmarksnumbered=false,bookmarksopen=false,
 breaklinks=false,pdfborder={0 0 0},pdfborderstyle={},backref=false,colorlinks=true]
 {hyperref}
\hypersetup{allcolors=blue}

\makeatletter
\renewcommand\NAT@citesuper[3]{\ifNAT@swa
\unskip\hspace{1\p@}\textsuperscript{[#1]}%
\if\relax#3\relax\else\ [#3]\fi\else [#1]\fi\endgroup}
\makeatother

\usepackage [english]{babel}
\usepackage [autostyle, english = american]{csquotes}
\MakeOuterQuote{"}

\begin{document}
\setcitestyle{super}
\title{The Gaussian Kicked Rotor:\linebreak Periodic forcing with finite-width pulses and the role of shifting the kick}

\author{Jonathan Berkheim, Shaked Levy, David J. Tannor}

\affiliation {\it Department of Chemical and Biological Physics, Weizmann Institute of Science, 76100, Rehovot, Israel}

%\date{\today}

\begin{abstract}  
The Kicked Rotor is perhaps the simplest physical model to illuminate the transition from regular to chaotic motion in classical mechanics. It is also widely applied as a model of light-matter interactions. In the conventional treatment, the infinitesimal width of each kick allows an immediate integration of the equations of motion. This in turn allows a full description of the dynamics via a discrete mapping, the Standard Map, if one looks at the dynamics only stroboscopically. It turns out that this model is only part of a much richer story if one accounts for finite temporal width of the kick. In this letter, we formulate a general model of finite-width periodic forcing and derive a continuous set of maps that depend on a parameter shift $\Delta$ that allows one to capture the motion in both the driven and kicked regimes. The fixed points and symmetry of the mapping are shown analytically and numerically to depend on the value of the shift parameter.
\end{abstract}

\maketitle 
\section{Introduction}
Regular and chaotic motion have been extensively studied in classical mechanics for over 100 years.\citep{Poincare1892LesCeleste,Einstein1917OnEpstein,Birkhoff1927DynamicalSystems} It was recognized that even in Hamiltonians with a single degree-of-freedom, which are integrable by construction, introducing an explicit time-dependence might lead to chaotic motion.\citep{Tzemos2024FormalSystems,Wiggins2003IntroductionChaos} Among such Hamiltonians, time-periodic kicked systems have been particularly well-studied.\citep{Fishman1982ChaosLocalization,Haake1987ClassicalTop,Moore1995AtomRotor,Artuso2011KickedModel}

A paradigm for the transition from regular to chaotic dynamics in time-periodic systems is provided by the Kicked Rotor,\citep{Casati1979StochasticSystems} a pendulum subjected to $\delta$-comb forcing in time:
\begin{equation}
\label{KR}
H(\theta,p,t)=\frac{p^2}{2}+K\cos\theta\sum_{n=-\infty}^{\infty}\delta(t-nT),
\end{equation}
where $\theta$ and $p$ are the angular position and momentum of the pendulum, respectively, and $K$ is the kick amplitude. For this Hamiltonian there exists a mapping that corresponds to a stroboscopic view of the dynamics immediately before each kick, the so-called Standard Map:\citep{Chirikov1969ResearchStochasticity}
\begin{equation}
\label{standard-map}
\begin{cases}
p_{n+1}=p_{n}+K\sin\theta_{n}\\
\theta_{n+1}=\theta_{n}+p_{n+1}
\end{cases}\hspace{0.5cm}.
\end{equation}
While the mapping provides only a stroboscopic view, its geometrical structure captures the dynamical properties of the entire phase space; it exhibits a transition from purely regular motion, to mixed motion, to purely chaotic motion, depending on the magnitude of $K$.\citep{Heller2018TheSpectroscopy,Lichtenberg1992RegularDynamics} Historically, the Standard Map actually preceded the Kicked Rotor.\citep{Chirikov2008ChirikovMap}

Although the Kicked Rotor is a very enlightening model, $\delta$ functions do not exist in nature. Pulses generated in atomic, molecular, and optical systems necessarily have a finite width in time, even if small. For instance, attosecond pulse trains, which are commonly utilized as a “camera” to probe ultrafast processes, are very narrow in time, but due to the cutoff frequency are not a $\delta$-comb in the strict mathematical sense. It is not obvious whether an analytical mapping exists for the extended case where the forcing has finite width.

In this letter, we formulate a general mapping for the case of finite-width periodic forcing. We derive a continuous set of maps that depend on a shift parameter $\Delta$ that allows one to capture the motion in both the driven (finite-width forcing) and kicked (instantaneous forcing) regimes. In addition to $n$, which specifies the discrete time increments for the dynamical variables $x(nT),p(nT)$, we introduce a parameter $\Delta$ such that the variables of the mapping are now $x(nT\pm\Delta),p(nT\pm\Delta)$. This allows for the sampling of the dynamics in the environment of each integer time period. For an ensemble of initial conditions, in order to correlate the results of the full dynamics with those of the mapping, a short-time propagation under the pulse envelope, dependent on $\Delta$, is required. Collecting the maps for all $n$ and $\Delta$ provides the continuous solution for Hamilton's EOMs. Recovering the continuous solution for Hamilton's EOMs from the discrete mappings is unusual and noteworthy. 

As seen below, there is excellent agreement between the $\Delta$-dependent mapping and the exact dynamics obtained via numerical integration. Significantly, the fixed points and symmetry of the mapping depend on the value of the shift parameter, as shown both analytically and numerically. It emerges that the Standard Map is only part of a much richer story if one accounts for finite temporal width of the kick.

\section{Theory and formulations}
\subsection{Generalized map of kicked dynamics}
\textit{Equations of motion.} Consider a general time-periodic Hamiltonian,
\begin{equation}
\label{general-H}
H(\textbf{x},\textbf{p},t)=\frac{\textbf{p}^2}{2}+V(\textbf{x})\sum_{n=-\infty}^{\infty}f(t-nT),
\end{equation}
where $F(t)\equiv\sum_{n=-\infty}^{\infty}f(t-nT)$ is a forcing function. The associated Hamilton’s EOMs are:
\begin{equation}
    \label{Hamilton-eq-gen}
\begin{cases}
\dot{\textbf{x}}=\textbf{p}\\
\dot{\textbf{p}}=-\nabla V(\textbf{x})\sum_{n=-\infty}^{\infty}f(t-nT)
\end{cases}\hspace{0.5cm}.
\end{equation}
We note several properties of this forcing which will guide our treatment below: (1) The forcing is periodic, i.e., $F(t)=F(t+T)$; (2) most of the sub-cycle motion is of a free particle, namely $\int_{-\infty}^{\infty}f(t)dt\approx\int_{nT-\varepsilon}^{nT+\varepsilon}f(t)dt$, where $\varepsilon\ll T$ is the decay time of $f(t)$; (3) the positions are frozen along the kick, i.e., $\{\textbf{x}(t)\hspace{0.1cm}|\hspace{0.1cm}t \in[nT-\varepsilon,nT+\varepsilon]\}=\text{const}$.\citep{R.Blumel1997ChaosPhysics} More rigorously, since $\textbf{x}(nT\pm\Delta)=\textbf{x}(nT)\pm\Delta\cdot \textbf{p}(nT)+\mathcal{O}(\Delta^2)$, for small enough $\Delta$ the first-order and higher-order terms are negligible, given that the momentum is finite; (4) we are interested in the stroboscopic dynamics. We thus define $(\textbf{x}_{n},\textbf{p}_{n}^{(\Delta)})\equiv(\textbf{x}(nT+\Delta),\textbf{p}(nT+\Delta))$, where $\Delta\in[nT-\varepsilon,nT+\varepsilon]$ denotes an arbitrary shift with respect to $nT$ along the kick, and $\textbf{x}_{n}$ is independent of $\Delta$ due to property (3).  

\begin{figure*}[t!]
    \centering
    \includegraphics[width=1\linewidth]{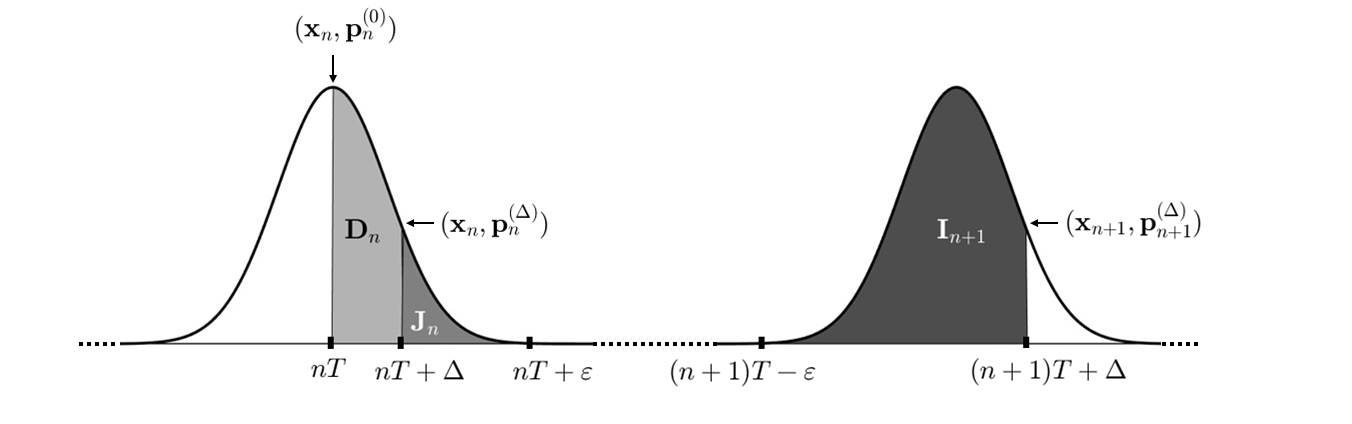}
    
    \caption{The impulse functions $\textbf{J}_{n}$ and $\textbf{I}_{n+1}$, as well as the drift $\textbf{D}_{n}$, depicted as fractional areas under the Gaussian kicks.}
    \label{fig:areas}
\end{figure*}

Using property (2), a formal integration of the momentum EOM in eqs. \ref{Hamilton-eq-gen} involves two impulse functions:
\begin{equation}
\label{impulse-J}
\textbf{J}_{n}^{(\Delta)}=-\int_{nT+\Delta}^{nT+\varepsilon}\nabla V(\textbf{x})f(t-nT)dt, 
\end{equation}
that is accumulated towards the decay of the kick, and
\begin{equation}
\label{impulse-I}
\textbf{I}_{n+1}^{(\Delta)}=-\int_{(n+1)T-\varepsilon}^{(n+1)T+\Delta}\nabla V(\textbf{x})f(t-nT)dt,
\end{equation}
that is accumulated from the ramp-up of the next kick. Fig. \ref{fig:areas} presents both impulse functions.

From property (3), it follows that the potential is frozen along the integration. Combining this with the periodicity property (1) we obtain: $\textbf{J}_{n}^{(\Delta)}=-\nabla V(\textbf{x}_{n})\int_{\Delta}^{\varepsilon}f(t)dt$ and $\textbf{I}_{n+1}^{(\Delta)}=-\nabla V(\textbf{x}_{n+1})\int_{-\varepsilon}^{\Delta}f(t)dt$, such that the impulses are the products of the fractional area under the kick and the local force.

\textit{Mapping.} In the same fashion as the construction of the Standard Map, we will use these impulse functions for the formulation of a mapping dependent on the discrete variable $n$. See fig. \ref{fig:areas} as a pictorial representation. As for the position, it follows from the position equation in eq. \ref{Hamilton-eq-gen} that:
\begin{equation}
\label{general-map-position}
\textbf{x}_{n+1}=\textbf{x}_{n}+\left(\textbf{p}_{n}^{(\Delta)}+\textbf{J}_{n}^{(\Delta)}\right)T, 
\end{equation}
where $\textbf{p}_{n}^{(\Delta)}+\textbf{J}_{n}^{(\Delta)}$ is the intermediate momentum which propagates the system as a free particle motion before the next kick (due to property (3), $\textbf{I}_{n+1}$ is excluded). As for the momentum, it follows from the momentum equation in eq.  \ref{Hamilton-eq-gen}:
\begin{equation}
\label{general-map-momentum}
\textbf{p}_{n+1}^{(\Delta)}=\textbf{p}_{n}^{(\Delta)}+\textbf{J}_{n}^{(\Delta)}+\textbf{I}_{n+1}^{(\Delta)}.
\end{equation}
The time interval between iterations $n$ and $n+1$ is always $T$, but in contrary to the Standard Map - where $n$ was recognized as a sole instant before (or after) the kick - here, one shall specify what exactly the time $n$ means with respect to the pulse shape; the parameter $\Delta$ dictates the exact moment when $\textbf{x}_{n},\textbf{p}_{n}$ are sampled with respect to the center of the pulse. Therefore, one shall select the value of $\Delta$ prior to the mapping, and to keep its value constant between iterations.

\textit{Reconstruction of continuous dynamics from mapping.} An ensemble of initial conditions which are properly selected to represent the same solution to the EOMs (see next paragraph) can be mapped with various values of $\Delta$, selected prior to the mapping. Then, the $n$-iteration can be collected with successive values of $\Delta$:
\begin{equation}
\label{continous-mapping}
\mathbb{M}=\left\{\begin{pmatrix}\textbf{x}(nT-\Delta_{N})\\\textbf{p}(nT-\Delta_{N})\end{pmatrix},...,\begin{pmatrix}\textbf{x}(nT)\\\textbf{p}(nT)\end{pmatrix},...,\begin{pmatrix}\textbf{x}(nT+\Delta_{N})\\\textbf{p}(nT+\Delta_{N})\end{pmatrix}\right\}.
\end{equation}
This set, $\mathbb{M}$, is a collection of positions and momenta sampled along the interval $[nT-\varepsilon,nT+\varepsilon]$. Once the set is defined for evenly-spaced values of $\Delta\in[-\varepsilon,\varepsilon]$ - such that $\Delta_{N}=\varepsilon$ - and the limit $N\to\infty$ is considered, then the continuous $\textbf{x}(t),\textbf{p}(t)$ are reconstructed under the pulse envelope. In other words, we solve the EOMs for the intervals $[nT-\varepsilon,nT+\varepsilon]$, for any $n$. The complementary intervals, i.e., $[nT+\varepsilon,(n+1)T-\varepsilon]$, correspond to free particle motion. See later discussion in Sec. III, as well as visual demonstration for the accuracy of these statements, in fig. \ref{fig:zoom_in_out}.

\textit{Initial conditions.} We now consider how to match the initial conditions of the dynamics with the initial conditions of the GKR maps. If the dynamics starts from $t=0$ with the initial conditions $(\mathbf{x}_{0},\mathbf{p}_{0})_{\text{d}}$, then the initial conditions of the map, $(\textbf{x}_{0},\textbf{p}_{0}^{(0)})_{\text{m}}$, where $\Delta=0$, remain the same: $(\textbf{x}_{0},\textbf{p}_{0}^{(0)})_{\text{m}}=(\mathbf{x}_{0},\mathbf{p}_{0})_{\text{d}}$. However, for any map with $\Delta\neq0$, $(\textbf{x}_{0},\textbf{p}_{0}^{(\Delta)})_{\text{m}}$ are actually $(\textbf{x}_{0},\textbf{p}_{0})_{\text{d}}$ that undergo a short-time propagation with duration $\Delta$, namely $(\textbf{x}_{0},\textbf{p}_{0}^{(\Delta)})_{\text{m}}=(\textbf{x}_{\Delta},\textbf{p}_{\Delta})_{\text{d}}$. Due to property (3), $\textbf{x}_{\Delta,\text{d}}=\textbf{x}_{0,\text{d}}$, while $\textbf{p}_{\Delta,\text{d}}=\textbf{p}_{0,\text{d}}+\textbf{D}_{0}^{(\Delta)}$, where we define an auxiliary function:
\begin{equation} 
\label{drift}
 {\textbf{D}}_{n}^{(\Delta)}\equiv -\nabla V(\textbf{x}_{n})\int_{0}^{\Delta} f(t)dt, 
 \end{equation}
as the accumulated impulse between the center of the kick and $\Delta$ (see fig. \ref{fig:areas}). All in all, this short-time propagation leads to the following shift of the initial conditions:
\begin{equation}
\label{IC}
\begin{pmatrix}
\textbf{x}_{0}\\
\textbf{p}_{0}^{(\Delta)}
\end{pmatrix}_{\text{m}}=
\begin{pmatrix}
\textbf{x}_{0}\\
\textbf{p}_{0}
\end{pmatrix}_{\text{d}}+\begin{pmatrix}
\textbf{0}\\
\textbf{D}_{0}^{(\Delta)}
\end{pmatrix},
\end{equation}
which is essential in order to obtain agreement between the dynamics and the mapping for different values of $\Delta$.

\subsection{Gaussian Kicked Rotor (GKR)}
We now consider a smoothed version of the Kicked Rotor, such that the forcing resembles a $\delta$-comb:
\begin{equation}
\label{gaussian-KR}
H(\theta,p,t)=\frac{p^2}{2}+K\cos\theta\sum_{n=-\infty}^{\infty}g(t-n),
\end{equation}
where $g(t)=\frac{1}{\sqrt{2\pi\sigma^2}}e^{-t^2/2\sigma^2}$. We choose $\varepsilon$ in properties (2) and (3) as $\varepsilon=3\sigma$, namely $g(\pm\varepsilon)=0.001g(0)\approx 0$); moreover, we take $T=1$ for simplicity. A $\delta$-comb can be recovered in the limit $\sigma\to0$. We note that one is free to choose any other function that approaches a $\delta$ function in a certain limit.\citep{Arfken2005MathamaticalPhysicists}

Henceforth, this model will be called the “Gaussian Kicked Rotor” (GKR). The associated Hamilton’s equations are:
\begin{equation}
    \label{Hamilton-eq-GKR}
\begin{cases}
\dot{\theta}=p\\
\dot{p}=K\sin\theta\sum_{n=-\infty}^{\infty}g(t-n)
\end{cases}\hspace{0.5cm}.
\end{equation}
Applying the abovementioned procedure (eqs. \ref{general-map-position}, \ref{general-map-momentum}), one obtains the following \textit{set of maps}: 
\begin{equation}
    \label{GKR-map}
\begin{cases}
\theta_{n+1}=\theta_{n}+p_{n}^{(\Delta)}+K\sin\theta_{n}\mathcal{A}^{(-\Delta)}\\
p_{n+1}^{(\Delta)}=p_{n}^{(\Delta)}+K\sin\theta_{n}\mathcal{A}^{(-\Delta)} +K\sin\theta_{n+1}\mathcal{A}^{(+\Delta)}
\end{cases},
\end{equation}
where $\mathcal{A}^{(\pm\Delta)}\equiv\frac{1}{2}\left[1\pm\text{erf}\left(\frac{\Delta}{\sqrt{2}\sigma}\right)\right]$. Note that for $\Delta=-\varepsilon$ we have  $\mathcal{A}^{(+\Delta)}|_{\Delta=-\varepsilon}\approx0$ and $\mathcal{A}^{(-\Delta)}|_{\Delta=-\varepsilon}\approx1$ and the Standard Map is recovered (see fig. \ref{fig:map_rotation_combined}). Thus, the Standard Map is just a single frame in the GKR set, the frame immediately before each kick. 
It is also clear that the GKR maps (for every $\Delta$) are area-preserving, just like the Standard Map.\citep{Lichtenberg1992RegularDynamics}
Since the maps are periodic in both $\theta$ and $p$, it is sufficient to consider $\theta$ and $p$ with modulo $2\pi$, such that the domain $[0, 2\pi]$ captures the whole stroboscopic dynamics in phase space.

\section{Analysis}
\subsection{Characterization of the GKR maps}
Our characterization of the GKR maps will focus on two properties: spatiotemporal symmetry and fixed points. Our main interest is in the dependence on $\Delta$, since this parameter determines the driven dynamics, which do not exist in the Standard Map.

\begin{figure}[ht!]
    \centering
    \includegraphics[width=0.85\linewidth]{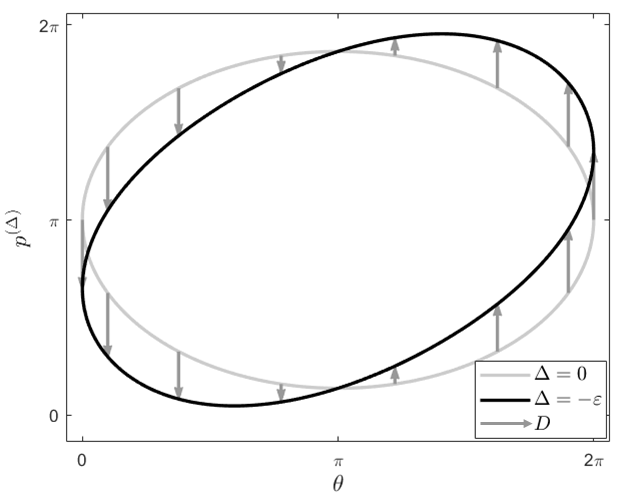}
    \caption{A representative regular trajectory in the GKR model, located around an elliptical fixed point, and how it undergoes a rotation in phase space due to maximal variation of $\Delta$.}
    \label{fig:ellipse}
\end{figure}

\textit{Spatiotemporal symmetry}. A prominent feature of the generalized map (eqs. \ref{general-map-position}, \ref{general-map-momentum}) is that for all reference points $({\textbf{x}}_{n},{\textbf{p}}_{n}^{(0)})$ generated by this map, a variation of $\Delta$ is manifested as a contribution to the momentum, ${\textbf{D}}_{n}^{(\Delta)}$, while ${\textbf{x}}_{n}$ remains frozen:
\begin{equation}
\label{symmetry}
\begin{pmatrix}
\textbf{x}_{n}\\
\textbf{p}_{n}^{(0)}
\end{pmatrix}\xrightarrow{\Delta}
\begin{pmatrix}
\textbf{x}_{n}\\
\textbf{p}_{n}^{(0)}+\textbf{{D}}_{n}^{(\Delta)}
\end{pmatrix}.
\end{equation}
Thus, each point in phase space experiences a drift only in the momentum direction with a magnitude proportional to $-\nabla V(\textbf{x})$ and an impulse determined by $\Delta$. Note that eq. \ref{symmetry} contains the short-time propagation of the initial conditions, as shown in eq. \ref{IC}.  Note further that a specific symmetry of $V(\textbf{x})$ will be expressed in the symmetry of the drift $\textbf{D}^{(\Delta)}$ with respect to $\textbf{x}$ axis (see. fig. \ref{fig:ellipse}). 

In the case of the GKR Hamiltonian (eq. \ref{gaussian-KR}),  $V(\theta)=K\cos\theta$. Substituting eq. \ref{gaussian-KR} into eq. \ref{symmetry} we find that $D_n^{(\Delta)}=K\sin \theta_{n} \text{erf}(\frac{\Delta}{\sqrt{2}\sigma})$ and thus:
\begin{equation}
\label{GKR-symmetry}
\begin{pmatrix}
\theta_{n}\\
p_{n}^{(0)}
\end{pmatrix}\xrightarrow{\Delta}
\begin{pmatrix}
\theta_{n}\\
p_{n}^{(0)}+K\sin\theta_{n}\text{erf}\left(\frac{\Delta}{\sqrt{2}\sigma}\right)
\end{pmatrix}.
\end{equation}
Note that since $D_n^{(\Delta)} \propto K\sin \theta_{n}$, for fixed $\Delta$ the drift exhibits symmetry in $\theta$. Dividing the interval $[0,2\pi]$ into four equally sized regions: $[0,\pi/2]$, $[\pi/2,\pi]$, $[\pi,3\pi/2]$, $[3\pi/2,2\pi]$ we label these regions A,B,C and D. Any point with $\theta_{n}$ in Region A drifts along the momentum axis with same magnitude $\left|D_{n}^{(\Delta)}\right|$ as the point in Region B with $\theta_{n}+\pi/2$. Any point  $\theta_{n}+3\pi/2$ in Region C drifts along the momentum axis with the same magnitude as a point in Region D with $\theta_n+\pi$,  but with drift  $-\left|D_{n}^{(\Delta)}\right|$, i.e.,  in the opposite direction from the corresponding points in Regions A and B; this is viewed all together as the rotation of the map (see fig. \ref{fig:ellipse}). Since $D^{(\Delta)}\propto\sin\theta$, the magnitude of the drift is nonuniform along $\theta$. As a function of $\Delta$, the drift is anti-symmetric and the map rotates in a nonlinear manner.

    \begin{figure*}[ht!]
    \centering
    \includegraphics[width=1\linewidth]{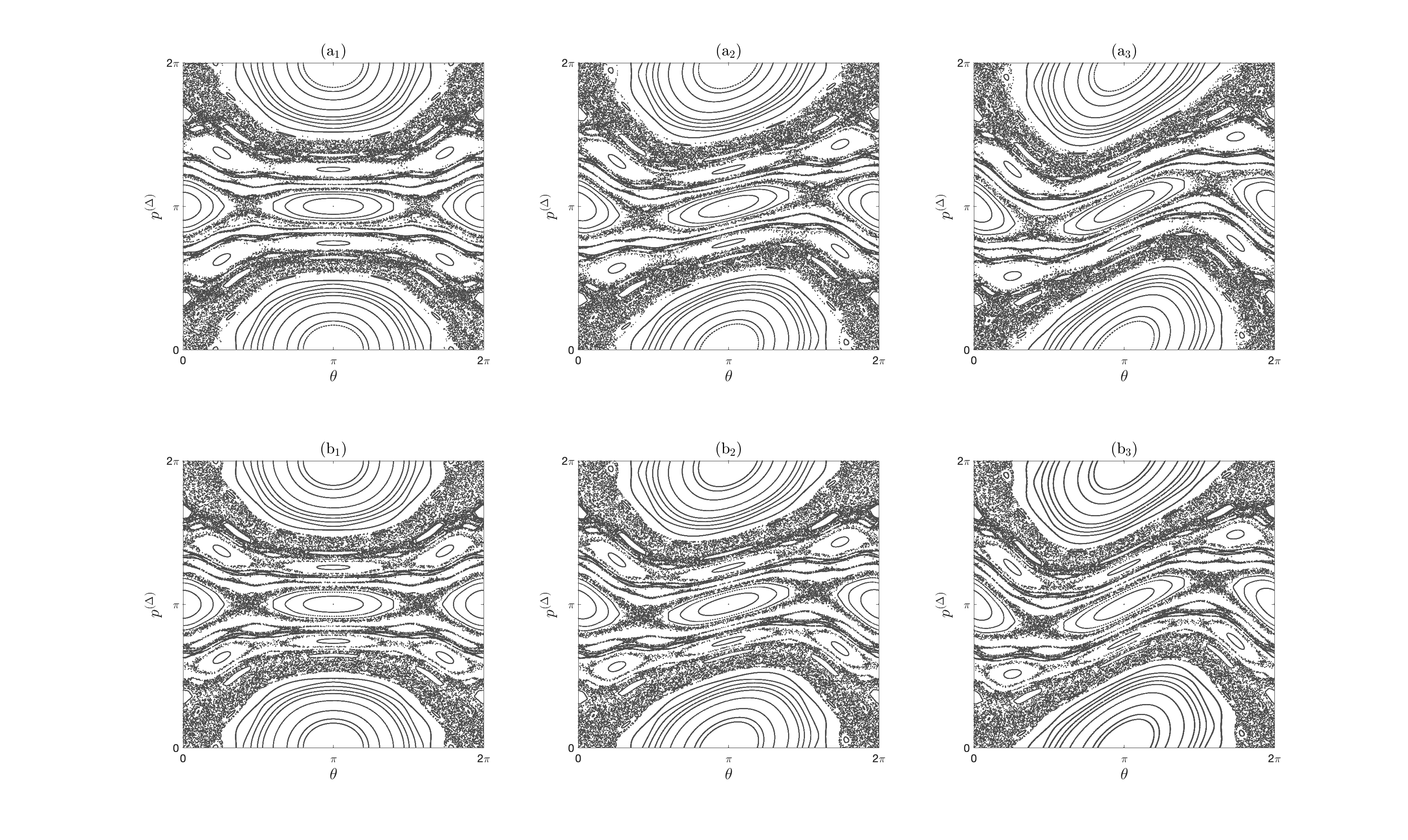}
    \caption{Representative maps of the GKR for varying $\Delta$: (1) $\Delta=0$, (2) $\Delta=-\frac{\varepsilon}{3\sqrt2}$, (3) $\Delta =-\varepsilon$. The upper plots are analytical (eqs. \ref{GKR-map}) and the lower plots are numerical. For the numerical RK$_\text{4}$ simulations we considered $\varepsilon=10^{-3}T$.}
    \label{fig:map_rotation_combined}
\end{figure*}

We suggest an additional point of view. Any EOM derived from a potential that is well-approximated by a simple harmonic oscillator around its minimum ($V(\textbf{x})\sim \textbf{x}^{2}$), and subjected to a time-periodic kick, can be cast in the form of Hill’s differential equation:\citep{Magnus1979HillsEquation}
\begin{equation}
\label{Hill}
\ddot{\textbf{x}}+\left[c_{0}+2\sum_{n=1}^{\infty}c_{n}\cos n\omega t\right]\textbf{x}=0,
\end{equation}
where $c_{n}$ are expansion coefficients. In particular, this is valid for regular trajectories of Kicked Rotor, where the selection $c_{n}=1,\hspace{0.2cm}\forall n\in \textbf{N}_{0}$ yields the $\delta$-comb. Since a Gaussian comb can be expanded with a proper selection of the $c_{n}$, the regular solutions of the GKR model, where effectively $V(\theta)\sim \theta^2$, will inherit the nature of Hill’s regular solutions.

A special case of Hill’s equation is the Mathieu equation, which corresponds to a time-periodic \textit{drive}, i.e., when $c_{0}=c_{1}=1$ and $c_{n}=0,\hspace{0.2cm} \forall n\geq 2$. It appears that regular solutions of this equation, which are analytically given by a linear combination of elliptical functions, exhibit a similar spatiotemporal symmetry along the drive: there is a drift that causes a rotation of phase space, which resembles the rotation in the GKR model, as in fig. \ref{fig:ellipse}. Therefore, this rotational behavior, which depicts the evolution of the phase space along the driven motion, is not unique for our model, but appears to be relevant to driven dynamics in general.

\textit{Fixed points}. Applying the conditions $\theta_{n+1}=\theta_{n}+2\pi m$ ($m\in \textbf{Z}$) and $p_{n+1}^{(\Delta)}=p_{n}$ on the angular part of eqs. \ref{GKR-map} yields:
\begin{equation}
\label{period-one}
\sin\theta^{\star}\left[\mathcal{A}^{(+\Delta)}+\mathcal{A}^{(-\Delta)}\right]=\sin\theta^{\star}=0,
\end{equation}
such that the period-one fixed points are $(\theta^{\star},p^{\star})=(0,0),(\pi,0)$, the same as in the Standard Map,\citep{Lichtenberg1992RegularDynamics} but now this is correct for \textit{every} $\Delta$.

Applying the conditions $\theta_{n+2}=\theta_{n}+2\pi m_{2}$ and $p_{n+2}^{(\Delta)}=p_{n}$ on the eqs. \ref{GKR-map}, and selecting the case $\theta_{n+1}=-\theta_{n}+2\pi m_{1}$ (where $m_{1},m_{2}\in \textbf{Z}$; see appendix) yields:
\begin{equation}
\label{period-two}
\begin{cases}
 K\sin\theta^{\star\star}+4\theta^{\star\star}=2\pi(2m_{1}-m_{2})\\
 p^{\star\star}(\theta^{\star\star})=\pi m_{2}+\frac{K}{2}\text{erf}\left(\frac{\Delta}{\sqrt{2}\sigma}\right)\sin\theta^{\star\star}
\end{cases}\hspace{0.5cm}.
\end{equation}
We solve these equations for two distinct selections of $m_{1},m_{2}$ to obtain a set of period-two fixed points, denoted $(\theta^{\star\star},p^{\star\star})$. These certain selections were chosen to emphasize the development of the period-two fixed points under variation of $\Delta$.

(i) For $2m_{1}=m_{2}$, we obtain the fixed point $(\theta^{\star\star},p^{\star\star})=(0,\pi)_{n}\rightleftarrows(\pi,\pi)_{n+1}$ that is not affected by any variation of $\Delta$; it can be also seen, by substituting this point’s foci $(n,n+1)$ in eq. \ref{GKR-symmetry}, that the point is invariant to the drift. 

(ii) For $2m_{1}=m_{2}+1$, we obtain a fixed point that its $\theta^{\star\star}$ is not affected by variation of $\Delta$, yet its $p^{\star\star}$ is affected. By solving eqs. \ref{period-two} for this condition, one obtains:
\begin{equation}
\label{fixed-point-sym}
 \small
\begin{pmatrix}
\theta^{\star\star}\\
\pi m_{2}+\frac{K}{2}\text{erf}\left(\frac{\Delta}{\sqrt{2}\sigma}\right)\sin\theta^{\star\star}
\end{pmatrix}_{n}\rightleftarrows 
\begin{pmatrix}
2\pi m_{2}-\theta^{\star\star}\\
\pi m_{2}-\frac{K}{2}\text{erf}\left(\frac{\Delta}{\sqrt{2}\sigma}\right)\sin\theta^{\star\star}
\end{pmatrix}_{n+1}.
\end{equation}
One can also predict the dependence this relation by substituting the fixed point with $\Delta=0$ in the eq. \ref{GKR-symmetry}, to obtain the LHS of eq. \ref{fixed-point-sym}; the resultant fixed points (for different $\Delta$) are mapped according to the symmetry rules described in the paragraph following  eq. \ref{fixed-point-sym}. All in all, a combination of drift and symmetry is equivalent to the mapping between the foci.

\begin{figure}[ht!]
    \centering
    \includegraphics[width=0.92\linewidth]{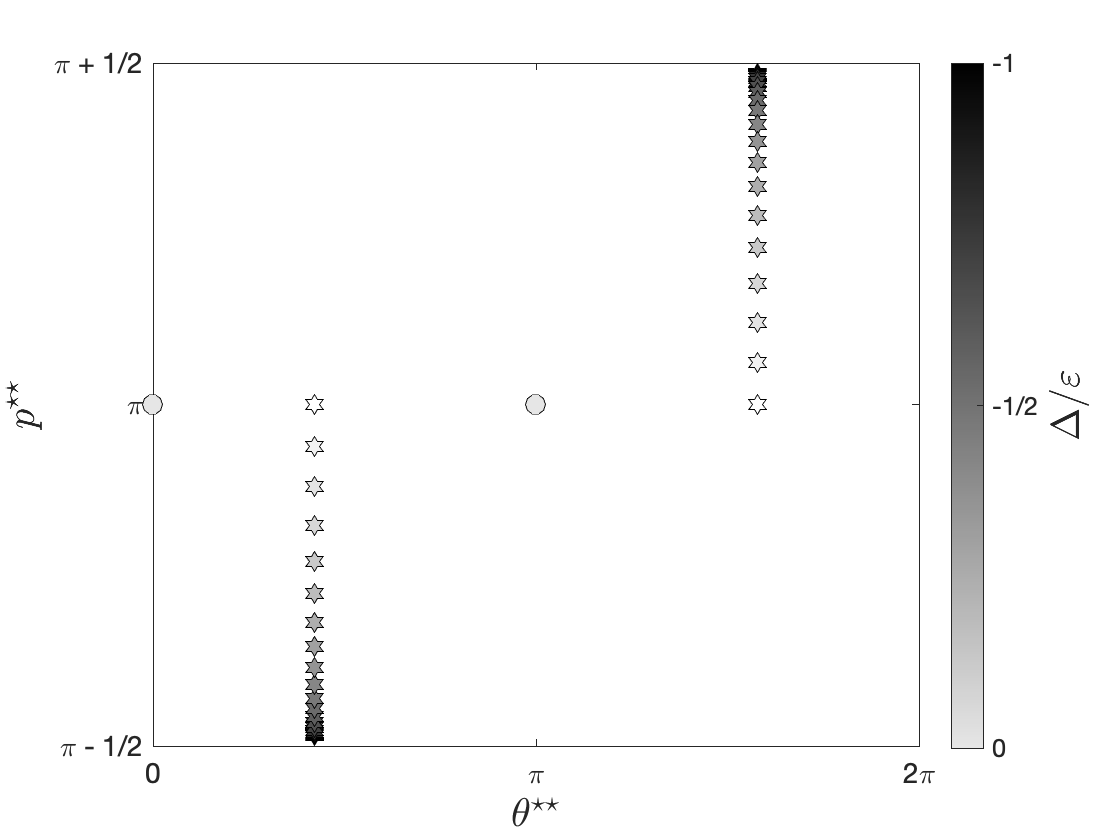}
    \caption{The “motion” of the period-two fixed points due to variation of $\Delta$. As seen, for $K=1$ the maximal absolute drift is $|D|=\left|\frac{K}{2}\sin\theta^{\star\star}\text{erf}\left(\frac{\varepsilon}{\sqrt{2}\sigma}\right)\right|=\frac{1}{2}$. The fixed points marked by circles emerge from $2m_{1}=m_{2}$ while those marked by stars emerge from $2m_{1}=m_{2}+1$.}
    \label{fig:fixed_points}
\end{figure}

\subsection{Comparison to numerical integration}
In order to evaluate the accuracy of the GKR maps with respect to the exact solutions of eqs. \ref{Hamilton-eq-GKR}, we numerically integrate the latter and compare their stroboscopic cuts to the GKR maps for different values of $\Delta$, as seen in fig. \ref{fig:map_rotation_combined}. The comparison indicates how accurately the map describes the dynamics arising from “realistic” kicks with finite width. 

We employed 4$^{\text{th}}$-order Runge-Kutta (RK$_4$) integration\citep{Butcher2000NumericalCentury} with identical domain of initial conditions as those examined in the GKR maps ($\delta\theta_{0}=\delta p_{0}=0.2\pi$). The trajectories evolve from $t=0$ to a fairly long time, $t=800T$, such that the initial conditions of the RK$_4$ dynamics are aligned with the center of the first Gaussian, and the corresponding initial conditions of the GKR maps are exactly as in eq. \ref{IC}. Convergence was verified with respect to the time step $\delta t$ and also with 6$^{\text{th}}$-order Runge-Kutta-Verner (RKV$_6$) integration, such that the RK$_4$ maps faithfully represent the solution to eqs. \ref{Hamilton-eq-GKR}.

The decay parameter $\varepsilon$ of each Gaussian kick is relatively small compared to $T$, such that properties (1-3) are obeyed to high fidelity, yet the integration is done with a small time step relative to the Gaussian width, $\varepsilon=10^{-3}$; we considered $\delta t=5\cdot 10^{-5}T$, such that the sampling of the forcing is fairly smooth. For a numerically converged solution, any differences between the GKR maps and the numerical solutions can stem from only two origins: (1) the slight variation of $\theta_{n}$ under the kick, and (2) the slight contribution to the momentum, as imposed by the tail of each Gaussian which exceeds the decay parameter $\varepsilon$. These differences are expected to appear only at long times of integration when the accumulated errors get considerable.

For all intents and purposes, the GKR and the RK$_4$ maps are identical for the same choice of $\Delta$: fixed points, main islands, island chains, separatrices, and chaotic layers. It is evident that the structure of the GKR maps under variation of $\Delta$ develops in the same way as in the RK$_{\text{4}}$ maps; this is true not only for specific values of $\Delta$ but for \textit{continuous} variation of $\Delta$, which follows perfectly the continuous variation of the RK$_\text{4}$ map. Therefore, the variation of $\Delta$ can be interpreted as the time evolution of the entire phase space under the kick, rather than just an “observational” parameter. This implies that the complete driven motion under the finite width pulse can be reconstructed as a collection of the GKR maps for continuously successive selections of $\Delta$.

    \begin{figure*}[ht!]
    \centering
    \includegraphics[width=1\linewidth]{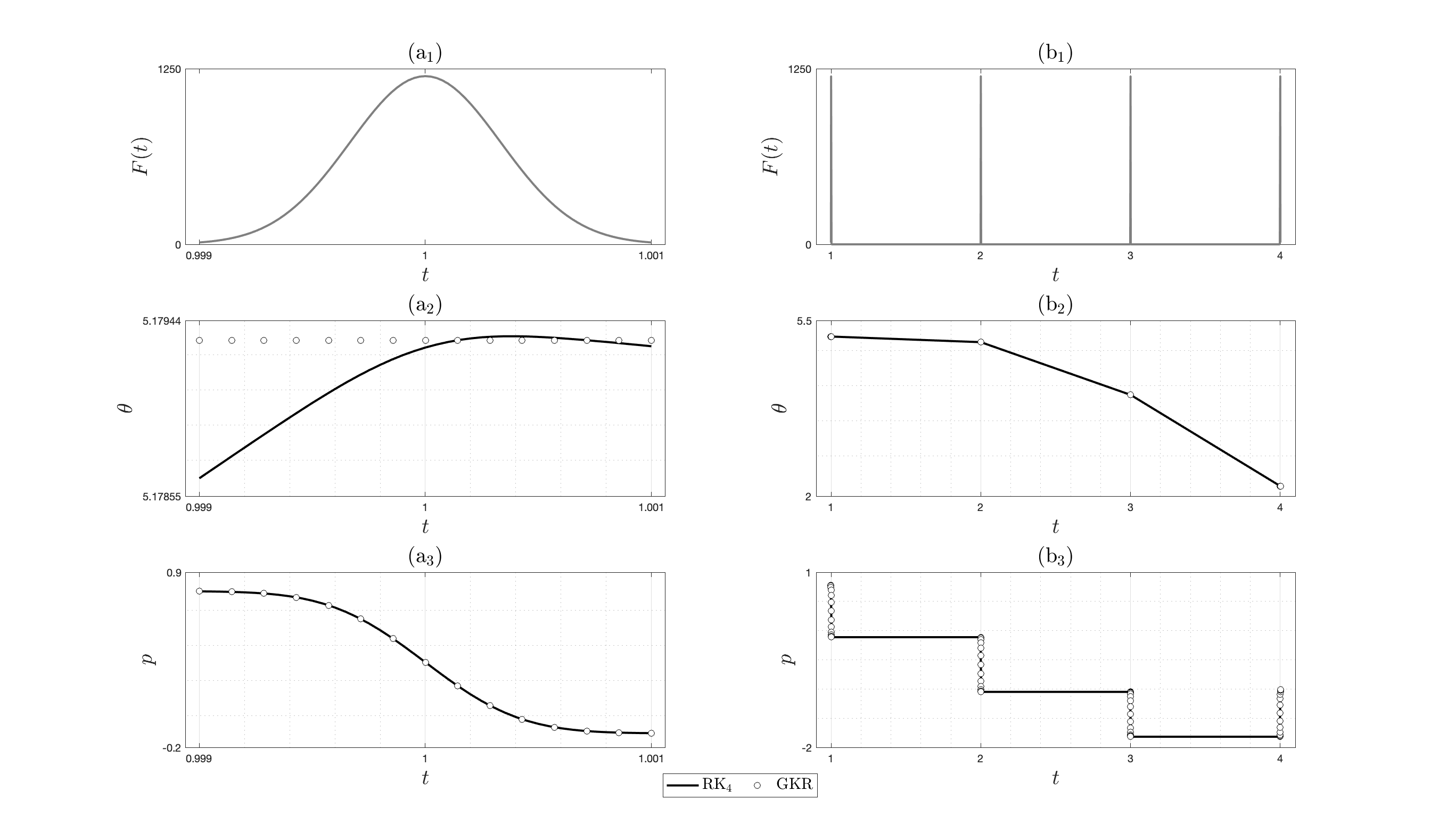}
    \caption{“Zoom-in” and “zoom-out” views over the evolution of certain regular trajectory, $\left(\theta_0,p_0^{(0)}\right)=(1.4\pi,0.4\pi)$, in the analytical and numerical sets, under (a) single kick and (b) sequence of few kicks: (1) the forcing itself, $F$, (2) the position, $\theta$, (3) the momentum, $p$.}
    \label{fig:zoom_in_out}
\end{figure*}

We demonstrate such a continuous collection for representative initial conditions in fig. \ref{fig:zoom_in_out}. The collection is compared to the complete trajectory obtained from RK$_{4}$, rather than a stroboscopic view as presented in fig. \ref{fig:map_rotation_combined}. The comparison is done on two temporal scales: column (a) shows the evolution under a single finite width pulse, and column (b) shows the evolution during several consecutive finite-width kicks, where most of the motion is that of a free particle. In (a), we see the meaning of $\Delta$ in the GKR maps as the timescale parameter that generates the driven dynamics. In (a$_2$), although $\theta^{(\Delta)}$ of the GKR is frozen during the kick, it is in very good agreement with the RK$_{\text{4}}$ value of $\theta(t)$; in (a$_3$), $p^{(\Delta)}$ is in virtually perfect agreement with the RK$_{\text{4}}$ value of $p(t)$. In (b$_1$), (b$_2$), the agreement of the GKR with the RK$_{\text{4}}$ solution is virtually perfect in both coordinate and momentum, in accordance with the stroboscopic maps. In summary, the frozen angle approximation barely affects the driven motion, all the more so in the “zoom-out” regime, where multiple kicks contribute.

\begin{figure}
    \centering
    \includegraphics[width=0.92\linewidth]{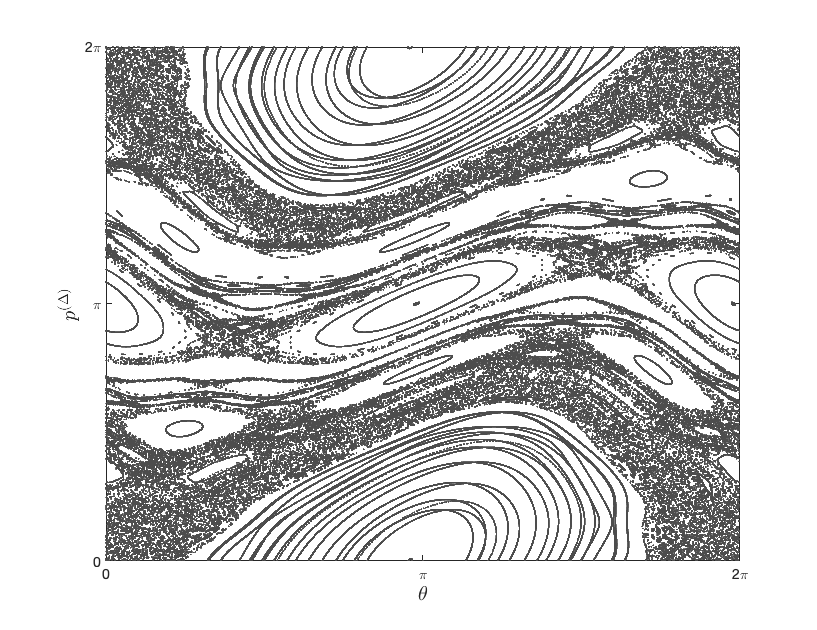}
    \caption{The phase portrait as in fig. \ref{fig:map_rotation_combined} ($\text{b}_3$), now with $\varepsilon=0.02$ ($>\varepsilon_c$). Note that several regular curves intersect each other, implying the breakdown of $2\pi$-periodicity in phase space.}
    \label{fig:above_critical}
\end{figure}

Clearly, at some point assumption (2) must break down. We find numerically that there is a critical value $\varepsilon_{\text{c}}=2\cdot10^{-3}$, above which the $2\pi$-periodicity in phase space is lost. This manifests itself in terms of regular curves that intersect each other. Fig. \ref{fig:above_critical} presents such a surface of section for $\varepsilon=10\varepsilon_{\text{c}}$, where the difference from the GKR maps in fig. \ref{fig:map_rotation_combined} is evident.

\section{Conclusion}
In this letter we formulated a mapping for periodic kicked dynamics with finite width pulses. The analytical solution was compared to numerical integration of the Kicked Rotor, where the kicks were of finite width and Gaussian shape.

We showed that in order to obtain agreement between the analytical mapping and the numerical solution, one must take into account the effect of the finite width pulse on the the initial conditions. Once this is done, the collection of mappings as a function of the continuous shift parameter $\Delta$ provides an excellent reconstruction of the full continuous dynamics under the kick. The dependence of the fixed points, as well as the symmetry of the surface of section as a function of $\Delta$ was analyzed both analytically and numerically. Both period-one and period-two fixed points are found. Interestingly, only half of the period-two fixed points change their position in phase space during the kick, while the others remain invariant.

Overall, there is excellent agreement between the analytical and numerical maps, such that the GKR maps indeed predict the actual solutions well. This agreement has two aspects: (1) the kicked dynamics, i.e., the stroboscopic phase space, which is a “zoom-out” of the trajectories, and (2) the driven dynamics, i.e., the “zoom-in” of the trajectories in the short-time interval between the tails of each kick. 

Several directions for future research come to mind. First, it would be interesting to apply the continuous mapping procedure for several other kicked systems in classical mechanics. Second, it would be interesting to examine the case where the pulse has a width larger than $\varepsilon_c$ and explore whether pulse width can have a qualitative effect on the transition between regular and chaotic motion. Third, it will be interesting to formulate a general solution for kicked-driven Hamiltonians in the quantum framework. 

\section{Appendix: calculation of fixed points}
For convenience, we will recall the GKR maps:
\begin{equation}
\label{app-1}
\begin{cases}
    p_{n+1}^{(\Delta)} = p_{n}^{(\Delta)} + K\sin\theta_{n} \mathcal{A}^{(-\Delta)} + K\sin\theta_{n+1} \mathcal{A}^{(+\Delta)} \\
    \theta_{n+1} = \theta_n + p_{n}^{(\Delta)} + K\sin\theta_{n} \mathcal{A}^{(-\Delta)}
    \end{cases}\hspace{0.5cm}.
\end{equation}
\subsection{Period-one fixed points}
In order to calculate the period-one fixed points, we will apply the conditions $\theta^{\star}\equiv\theta_{n+1}=\theta_{n}+2\pi m$ and $p^{\star}\equiv p_{n+1}^{(\Delta)}{(\Delta)}=p_{n}^{(\Delta)}$ on eqs. \ref{app-1}. From the momentum equation it arises that:
\begin{equation}
K\sin\theta^{\star}\mathcal{A}^{(-\Delta)}+K\sin\theta^{\star}\mathcal{A}^{(+\Delta)}=0,
\end{equation}
such that:
\begin{equation}
\sin\theta^{\star}\left[\mathcal{A}^{(+\Delta)}+\mathcal{A}^{(-\Delta)}\right]=\sin\theta^{\star}=0,
\end{equation}
and the solutions are $\theta^{\star}=0,\pi$ in the modulo-$2\pi$ domain. By setting these $\theta^{\star}$ in the angular equation we obtain that $p^{\star}(\theta^{\star})=0$. All in all, the period-one fixed points are $(\theta^{\star},p^{\star})=(0,0),(\pi,0)$.

\subsection{Period-two fixed points}
In order to calculate the period-two fixed points, we will first derive $\theta_{n+2},p_{n+2}^{(\Delta)}$ in terms of $\theta_{n+1},p_{n+1}^{(\Delta)}$:
\begin{equation}
\label{app-2}
\begin{cases}
    p_{n+2}^{(\Delta)} = p_{n+1}^{(\Delta)} + K\sin\theta_{n+1} \mathcal{A}^{(-\Delta)} + K\sin\theta_{n+2} \mathcal{A}^{(+\Delta)} \\
    \theta_{n+2} = \theta_{n+1} + p_{n+1}^{(\Delta)} + K\sin\theta_{n+1} \mathcal{A}^{(-\Delta)}
    \end{cases}\hspace{0.5cm}.
\end{equation}
Substituting eqs. \ref{app-2} into eqs. \ref{app-1} yields that:
\begin{widetext}
\begin{equation}
\label{app-3}
\begin{cases}
    p_{n+2}^{(\Delta)} = p_{n}^{(\Delta)} + K\sin\theta_{n} \mathcal{A}^{(-\Delta)} + K\sin\theta_{n+1} \mathcal{A}^{(+\Delta)} + K\sin\theta_{n+1} \mathcal{A}^{(-\Delta)} + K\sin\theta_{n+2} \mathcal{A}^{(+\Delta)}\\
    \theta_{n+2} = \theta_n + p_{n}^{(\Delta)} + K\sin\theta_{n} \mathcal{A}^{(-\Delta)} + p_{n}^{(\Delta)} + K\sin\theta_{n} \mathcal{A}^{(-\Delta)} + K\sin\theta_{n+1} \mathcal{A}^{(+\Delta)}
    \end{cases}\hspace{0.5cm},
\end{equation}
\text{which, within several algebraic steps, has the following form:}
\begin{equation}
\label{app-4}
\begin{cases}
    p_{n+2}^{(\Delta)}= p_{n}^{(\Delta)} + K\sin\theta_{n} \mathcal{A}^{(-\Delta)} + K\sin\theta_{n+1} \left[{A}^{(+\Delta)} + \mathcal{A}^{(-\Delta)}\right] + K\sin\theta_{n+2} \mathcal{A}^{(+\Delta)} \\
    \theta_{n+2}= \theta_n + 2p_{n}^{(\Delta)} + 2K\sin\theta_{n} \mathcal{A}^{(-\Delta)} + K\sin\theta_{n+1} \left[{A}^{(+\Delta)} + \mathcal{A}^{(-\Delta)}\right]
    \end{cases}\hspace{0.5cm},
\end{equation}

\text{and, after further rearrangement:}
\begin{equation}
\label{app-5}
\begin{cases}
    p_{n+2}^{(\Delta)}= p_{n}^{(\Delta)} + K\sin\theta_{n} \mathcal{A}^{(-\Delta)} + K\sin\theta_{n+1} + K\sin\theta_{n+2} \mathcal{A}^{(+\Delta)} \\
    \theta_{n+2}= \theta_n + 2p_{n}^{(\Delta)} + 2K\sin\theta_{n} \mathcal{A}^{(-\Delta)} + K\sin\theta_{n+1}
    \end{cases}\hspace{0.5cm}.
\end{equation}
\end{widetext}
The pair of conditions for period-two fixed points are:
\begin{equation}
\label{app-6}
\begin{cases}
    \text{(a)}  \quad \theta_{n+2} = \theta_n + 2\pi m_2  &(m_{2}\in\mathbf Z)\\
    \text{(b)}  \quad p_{n+2}^{(\Delta)} = p_{n}^{(\Delta)}
    \end{cases}\hspace{0.5cm},
\end{equation}
and substituting both of them in eqs. \ref{app-5} leads to:
\begin{equation}
\label{app-7}
    0 =  K\sin\theta_{n} \mathcal{A}^{(-\Delta)} + K\sin\theta_{n+1} + K\sin\theta_{n} \mathcal{A}^{(+\Delta)} ,
\end{equation}
which reduces to:
\begin{equation}
\label{app-8}
    \sin\theta_{n} + \sin\theta_{n+1} = 0.
\end{equation}
This condition breaks down into two cases:
\begin{equation}
\label{app-9}
\begin{cases}
    (\text{i}) & \theta_{n+1} = \pi - \theta_n + 2\pi m_1 \\
   (\text{ii}) & \theta_{n+1} = -\theta_n + 2\pi m_1
    \end{cases} \hspace{0.5cm} (m_{1}\in \mathbf{Z}),
\end{equation}
\begin{equation}
\label{app-10}
    \theta_{n+2} = \theta_n + 2p_{n}^{(\Delta)} + 2K\sin\theta_{n} \mathcal{A}^{(-\Delta)} + K\sin(-\theta_n + 2\pi m_1),
\end{equation}
such that we obtain:
\begin{equation}
\label{app-11}
    \theta_{n+2} = \theta_n + 2p_{n}^{(\Delta)} + 2K\sin\theta_{n} \mathcal{A}^{(-\Delta)} - K\sin\theta_{n}.
    \end{equation}
Substituting condition (a) into eq. \label{app-11} leads to:
\begin{equation}
    \label{app-12}
    \theta_n + 2\pi m_2 = \theta_n + 2p_{n}^{(\Delta)} + 2K\sin\theta_{n} \mathcal{A}^{(-\Delta)} - K\sin\theta_{n},
\end{equation}
which, after brief rearrangement reads as:
\begin{equation}
\small
\label{app-13}
    p_{n}^{(\Delta)} = \pi m_2 - K\sin\theta_{n} \left[\mathcal{A}^{(-\Delta)} - \frac{1}{2}\right]=\pi m_2 +\frac{K}{2}\text{erf}\left(\frac{\Delta}{\sqrt{2}\sigma}\right)\sin\theta_{n} .
    \end{equation}
Now, we set (ii) in the momentum equation among eqs. \ref{app-1}:
\begin{equation}
\label{app-14}
    -\theta_n + 2\pi m_1 = \theta_n + p_{n}^{(\Delta)} + K\sin\theta_{n} \mathcal{A}^{(-\Delta)},
\end{equation}
then combine eq. \ref{app-14} with eq. \ref{app-13} to get:
\begin{equation}
\small
\label{app-15}
    -2\theta_n + 2\pi m_1 = \pi m_2 - K\sin\theta_{n} \left[\mathcal{A}^{(-\Delta)} - \frac{1}{2}\right] + K\sin\theta_{n} \mathcal{A}^{(-\Delta)}.
\end{equation}
Eventually, another simplifications lead to:
\begin{equation}
\label{app-16}
    K\sin\theta_{n} + 4\theta_n  = 2\pi (2m_1 - m_2),
\end{equation}
which is the first equation in eqs. \ref{period-two} with the notation $\theta^{\star\star}=\theta_{n}$. The second equation in eqs. \ref{period-two}, on $p^{\star\star}$, pops up once we set the resultant $\theta^{\star\star}$ in eq. \ref{app-13}.
\bibliographystyle{apsrev4-2}
\bibliography{main}

\end{document}